\documentstyle[psfig,proceedings,bibstyle]{crckapb}
\newcommand{\mm}{$\lambda$ 1.2\,mm }

\newcommand{\hi}{H{\sc i}}
\begin{opening}
\title{Mapping the cold dust in edge-on galaxies \protect\\
 at 1.2\,mm wavelength}
\author{Nikolaus Neininger}
\institute{IRAM, F-38406 St.Martin d'H\`eres, France {\em and}\\
MPIfR, D-53121 Bonn, Germany}
\author{Michel Guelin}
\institute{IRAM, F-38406 St.Martin d'H\`eres, France}
\end{opening}
\begin{document}
\begin{abstract}

Using the IRAM 30-m telescope, we have mapped the $\lambda$ 1.2\,mm
continuum emission in the edge-on spiral galaxies NGC\,891, NGC\,5907
and NGC\,4565.  Generally, the $\lambda$ 1.2\,mm continuum correlates
remarkably well with the CO emission; the correlation with \hi\ is
however different for the observed galaxies: in NGC\,891, there is no
obvious correlation; in NGC\,5907 the continuum emission is extending
a bit further out than the CO and seems to be correlated with \hi\
peaks.

In NGC\,4565, however, the dust emission not only shows a central peak
and an inner ring like the CO, but also, like \hi{}, a weaker,
extended plateau. Comparable to the \hi , the 1.2 mm contours are
warped near the NW edge of the galaxy.

The average dust temperature in this galaxy is 18\,K near the center
and 15\,K in the \hi\ plateau. From the 1.2 mm continuum intensity and
the \hi\ line integrated intensity, we derive a dust absorption cross
section per H atom $\sigma{\rm _{1.2mm}^H}=5\times 10^{-27}$ cm$^2$ in
the plateau. This value is very close to that predicted for the local
diffuse clouds.

\end{abstract}
\section{Motivation of the study} 

Cold dust represents most of the interstellar dust in normal galaxies
and may be used as a tracer of both molecular and atomic gas (see Cox
\& Mezger 1989 and references therein).  Gu\'elin et al.\ (1993, 1995)
mapped the \mm continuum emission of two nearby spirals, NGC\,891 and
M\,51. This emission there was found to correlate tightly with CO and
poorly with \hi{}. It was not even clearly detected beyond NGC\,891's
molecular `ring', in a region where \hi\ emission is still strong. The
mean dust temperatures derived from the \mm and FIR flux densities
were found to be $\leq 20$\,K.
 
In order to further study the properties of the ISM, we observed two
more edge-on galaxies of similar type, NGC\,4565 (Neininger et al.\
1996) and NGC\,5907 (Dumke et al.\ 1996). In particular, NGC\,4565 was
chosen because of its weak CO emission: the dust emissivity per H atom
is on the average 2 -- 4 times larger for the molecular clouds than
for the \hi\ clouds. Thus, the dust associated with the atomic gas
becomes predominant only when the H$_2$ column density becomes very
small -- which is the case in the outer parts of NGC\,4565. The
edge-on geometry ($i \ge 85^{\circ}$) ensures long lines of sight in
the disk which helps to detect weak components of the ISM.

\begin{figure}
\vspace{-2.2cm}
\hbox{
\hspace{-2mm}\psfig{file=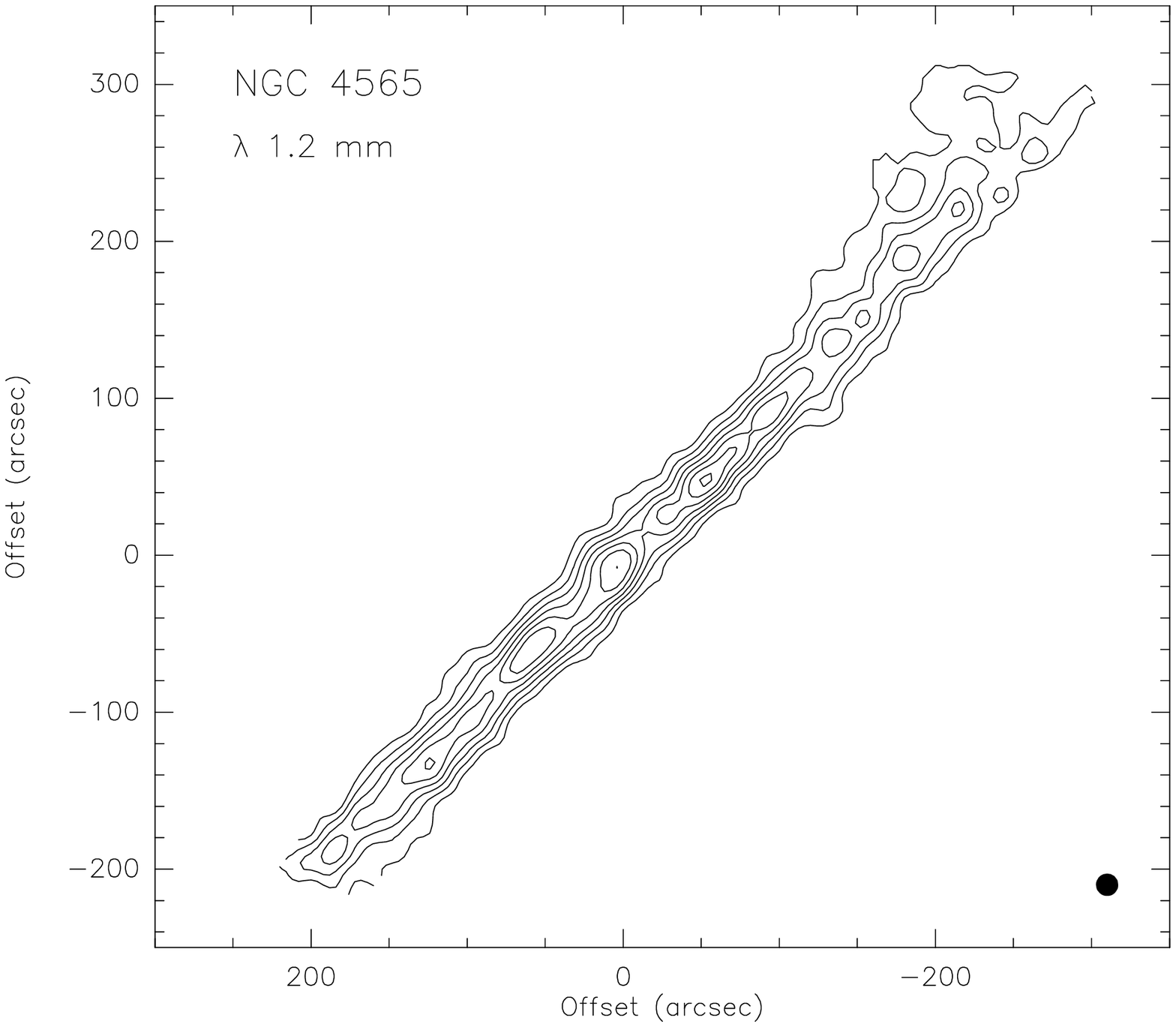,height=95mm}
\hspace{-5mm}\psfig{file=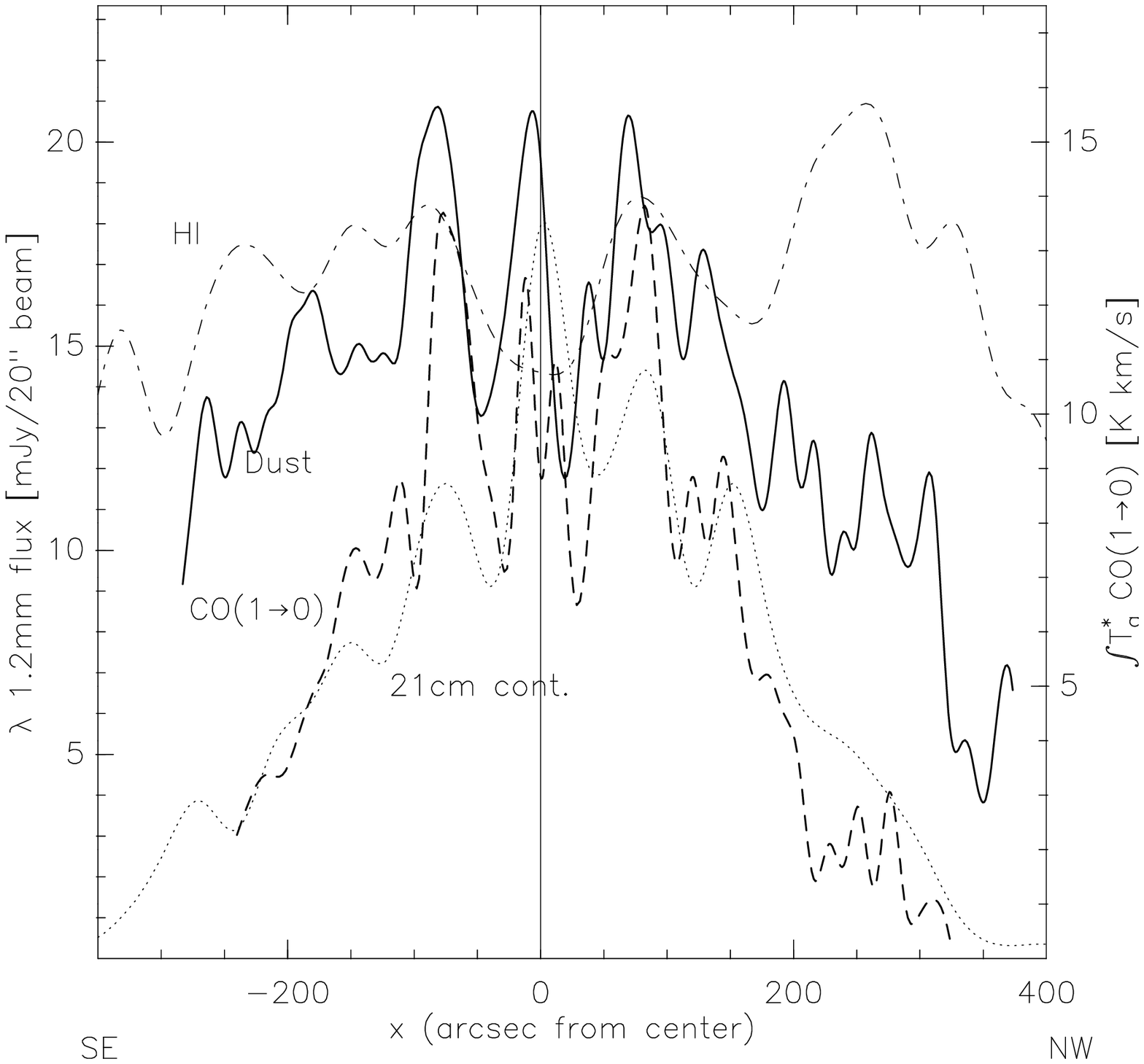,height=87mm,rheight=94mm}
}
\caption{a) The \mm continuum emission of NGC\,4565 smoothed to a resolution 
of $20''$; contour levels are 3, 6, ... 21 mJy/beam. b) The brightness
distributions of the integrated line intensities of atomic and
molecular gas along the major axis of NGC\,4565 together with the \mm
intensity and the 21-cm continuum.}
\end{figure}

\section{Observations}

The observations of NGC\,891 were carried out in February 1993 with a
7-channel bolometer array (Kreysa 1992), those of NGC\,5907 and
NGC\,4565 with an instrument upgraded to 19 channels in March 1995.
The beamwidths are $11''$ (HPBW), and the spacing between two adjacent
channels 20$''$. The equivalent bandwidth $\Delta_\nu$ and central
frequency $\nu_\circ$ should be close to 70\,GHz and 245\,GHz
(1.2\,mm), respectively (see Gu\'elin et al.\ 1995).

The maps are mosaics of up to 26 overlapping submaps, each of the size
of a few arcmin$^2$. During the observations, the subreflector was
wobbled at 2\,Hz in azimuth with a beam throw between $30''$ and
$56''$.  For each channel, a second order baseline was fitted to every
azimuth scan, the scans were combined and restored into single beam
maps and regridded in equatorial coordinates; after correcting the
intensities for atmospheric absorption, they were calibrated with
respect to the planets. Finally, the different channel maps were
combined to yield a single map which reached a maximum sensitivity of
$1\ldots 1.5$\,mJy per 12$''$ beam for the observations of NGC\,4565.

\section{Findings}

In all three objects, the bulk of the molecular gas lies within a
radius of 4--5 kpc from the center. The central component is
relatively bright in NGC\,891 and strongly dominating in NGC\,5907.
The central 3\,kpc region of NGC\,4565 hosts little interstellar
matter, except for a compact `ring' of molecular gas (of diameter
$\simeq1$ kpc).  In NGC\,891 and NGC\,4565 a strong molecular ring is
visible which contains most of the molecular gas. The bulk of the
atomic gas is situated further out, forming a broad `plateau' peaking
at $R\simeq 9-15$ kpc and extending up to $R \simeq 15-20$ kpc. The
outer plateau and molecular ring of NGC\,4565 show narrow density
structures, which are probably spiral arms.

The \mm emission, which is the most reliable tracer of interstellar
dust, follows closely the CO brightness distribution in the central
region and in the molecular ring; this holds for all three galaxies.
In NGC\,891 there is no sign of \mm emission further out (Gu\'elin et
al.\ 1993); in NGC\,5907, this emission is a bit more extended than
the CO disk and two bumps at a radius of about 10\,kpc coincide with
the \hi\ at a place where no CO is left (Dumke et al.\ 1996).  In
NGC\,4565, however, the \mm emission follows \hi\ in the outer regions
of the disk as soon as the CO emission becomes weak. This way, it
extends significantly further out than the emission of the molecular
gas and it is possible to derive the properties of the cold dust that
is associated with the atomic gas.

The outer \hi\ disks of all three galaxies are warped. Because of the
lack of detected emission, no trace of it can be seen in the \mm maps
of NGC\,891 and NGC\,5907. However, the onset of NGC\,4565's warp is
clearly visible in the 1.2\,mm cold dust emission at the NW side. In
the SE, the \hi\ warp is by far less prominent, but a hint at it is
seen just at the edge of our map.

\section{Results}

The most interesting result is the observation of \mm cold dust
emission at galactocentric distances $> 10$ kpc in the \hi\ ring and
in the warp of NGC\,4565. In these outer regions, where the gas is
mostly in the atomic form (this is indicated by the weakness of the CO
emission), it is possible to measure the dust emissivity per H-atom.
For the average dust tem\-pera\-ture of 15\,K (derived from a fit to
160$\,\mu$m, 200\,$\mu$m and 1.2\,mm data) the comparison of the
integrated \hi\ line intensity with the observed \mm flux intensity
yields an absorption cross section per H-atom $\sigma_\lambda^{\rm H}=
5\times10^{-27}$ cm$^2$\,(H-atom)$^{-1}$ and a mean dust absorption
coefficient $\kappa = 0.002\,{\rm cm^2g^{-1}}$ (Neininger et al.\
1996).  These cross sections and tem\-pera\-tures are similar to those
predicted in local diffuse clouds (see Draine \& Lee 1984).

The comparison of the three galaxies does not hint at a correspondence
between the Hubble type of the galaxy and the relative extent of the
cold dust emission; a trend is however visible for the distribution of
the molecular gas: there is a stronger ring in the galaxy of earlier
type and a stronger central component in the later type. NGC\,4565 is
possibly a barred galaxy (it has a peanut-shaped bulge and also the
non-circular motions observed in CO hint at this -- see Neininger et
al.\ 1996); this could influence the distribution of its ISM.
Observations at mm and FIR wavelengths are in progress to further
study these aspects.

The mass of the molecular gas in NGC\,4565 is low compared to the mass
of atomic gas. The molecular gas mass, inside the strip along the
major axis covered by our CO observations, is found to be $1.0\times
10^9$ M$_{\odot}$, when using the Galactic CO-to-H$_2$ conversion
factor ($X= 2.3\times10^{20}$cm$^{-2}{\rm K^{-1}km^{-1}s}$ -- Strong
et al.\ 1988), and $\simeq 0.4\times10^9$ M$_{\odot}$ when using the
values derived from the \mm emission. This corresponds respectively to
$\simeq$ 1/2 and 1/5 of the \hi\ mass in the same area ($2 \times10^9$
M$_{\odot}$).

\end{document}